\documentclass[conference]{IEEEtran}
\IEEEoverridecommandlockouts
\usepackage{cite}
\usepackage{amsmath,amssymb,amsfonts}
\usepackage{algorithmic}
\usepackage{graphicx}
\usepackage{textcomp}
\usepackage{xcolor}
\def\BibTeX{{\rm B\kern-.05em{\sc i\kern-.025em b}\kern-.08em
		T\kern-.1667em\lower.7ex\hbox{E}\kern-.125emX}}

\DeclareRobustCommand*{\IEEEauthorrefmark}[1]{%
	\raisebox{0pt}[0pt][0pt]{\textsuperscript{\footnotesize\ensuremath{#1}}}}

\begin{document}
\title{A Novel Massive MIMO Beam Domain\\ Channel Model}
\author{
	\IEEEauthorblockN{	Fan Lai\IEEEauthorrefmark{1,2},
		%Carlos F. L\'opez\IEEEauthorrefmark{2},
		Cheng-Xiang Wang\IEEEauthorrefmark{1,2,^*},
		Jie Huang\IEEEauthorrefmark{1,2},
		Xiqi Gao\IEEEauthorrefmark{1,2},
		and Fu-Chun Zheng\IEEEauthorrefmark{1,2}}	
	\IEEEauthorblockA{\IEEEauthorrefmark{1}\small{National Mobile Communications Research Laboratory,}}
	\IEEEauthorblockA{\small{School of Information Science and Engineering, Southeast University, Nanjing 210096, China.}}
	\IEEEauthorblockA{\IEEEauthorrefmark{2}\small{Purple Mountain Laboratories, Nanjing 211111, China.}}
	\IEEEauthorblockA{\IEEEauthorrefmark{*}\small{Corresponding Author: Cheng-Xiang Wang}}
	Email: \{lai\_fan, chxwang, j\_huang, xqgao\}@seu.edu.cn, fzheng@ieee.org}
\maketitle
\begin{abstract}
	A novel beam domain channel model (BDCM) for massive multiple-input multiple-output (MIMO) communication systems has been proposed in this paper. The near-field effect and spherical wavefront are firstly assumed in the proposed model, which is different from the conventional BDCM for MIMO based on the far-field effect and plane wavefront assumption. The proposed novel BDCM is the transformation of an existing geometry-based stochastic model (GBSM) from the antenna domain into beam domain. The space-time non-stationarity is also modeled in the novel BDCM. Moreover, the comparison of computational complexity for both models is studied. Based on the numerical analysis, comparison of cluster-level statistical properties between the proposed BDCM and existing GBSM has shown that there exists little difference in the space, time, and frequency correlation properties for two models. Also, based on the simulation, coherence bandwidths of the two models in different scenarios are almost the same. The computational complexity of the novel BDCM is much lower than the existing GBSM. It can be observed that the proposed novel BDCM has similar statistical properties to the existing GBSM at the cluster-level. The proposed BDCM has less complexity and is therefore more convenient for information theory and signal processing research than the conventional GBSMs.
\end{abstract}


\section{Introduction}
In wireless communication systems, MIMO technology can greatly enhance the channel capacity \cite{a1} and improve the reliability of communication systems without requiring additional bandwidth and power \cite{b1}. In recent years, massive MIMO technology has become popular because it can simplify the required signal processing\cite{b2} and provide a huge improvement in system performance over traditional MIMO technologies \cite{b3}.

The wireless channel model is a key metric to analyze the performance of the whole communication systems \cite{a2}. Recently, multiple GBSMs have been proposed and reviewed in \cite{a3}. Among them, a two-dimensional (2D) non-stationary wideband channel model was proposed in \cite{b4}, where all scatterers were located on multi-confocal ellipses. It considered the near-field effect and the non-stationarity of massive MIMO channel. In \cite{b5}, a three-dimensional (3D) non-stationary wideband massive MIMO channel model was proposed, which utilized a virtual link to represent the complicated scattering environment. In \cite{b6}, the authors proposed a general 3D GBSM for fifth generation (5G) wireless communication systems. It is capable of simulating massive MIMO, millimeter wave (mmWave), vehicle-to-vehicle (V2V), and high-speed train (HST) small-scale fading channels.

However, the complexity of the above-mentioned models is relatively high. Simplified GBSMs or correlation-based stochastic models (CBSMs) may be more applicable for information theory and signal processing research. CBSMs can be divided into classical independent and identically distributed (i.i.d.) Rayleigh fading channel model and correlated channel model \cite{b7}. Correlated channel models include the Kronecker-based stochastic model (KBSM) \cite{b8}, Weichselberger model \cite{b9}, and virtual channel representation (VCR) \cite{b10}. However, these conventional CBSMs did not take the near-field effect and non-stationarity into account, and thus are not suitable for massive MIMO channel modeling.
	
Recently, the BDCM was first proposed in \cite{b11}, which transferred the channel model into the angular domain or called beam domain with the far-field assumption. The conventional BDCM was applied to the research of information theory and signal processing, especially in the recent proposed beam division multiple access (BDMA) \cite{b11}. However, the conventional BDCM also did not consider the massive MIMO characteristics. A more accurate massive MIMO channel model which is convenient for information theory and signal processing research is therefore needed urgently.

In this paper, we propose a novel BDCM for massive MIMO systems and the major \textbf{contributions} are summarized as follows:
\begin{enumerate}
	\item The proposed novel BDCM is distinguished from the conventional BDCM by considering the near-field effect, spherical wavefront and space-time non-stationarity for the first time.
	\item The novel BDCM transfers an existing GBSM into the beam domain and according to the simulation results, the novel BDCM has the similar statistical properties with the existing GBSM in the cluster-level.
	\item By comparing the computational complexity of both models, it is found that the novel BDCM has the lower computational complexity and is more suitable for the research of signal processing and information theory than conventional GBSMs.
\end{enumerate}

The remainder of this paper is organized as follows. An existing 2D non-stationary wideband GBSM is presented in Section II. In section III, the proposed novel BDCM is discussed in details. Section IV shows the statistical property analysis of both two models. Results and discussions are presented in Section V. Conclusions are drawn in Section VI.

\section{An Existing 2D Non-stationary Wideband GBSM}
%\begin{figure}[tb]
%	\centerline{\includegraphics[width=0.4\textwidth]{Fig1_EllipseModel.pdf}}
%	\caption{Illustration of the existing 2D non-stationary wideband GBSM \cite{b4}.}
%	\label{fig}
%\end{figure}
%\subsection{Channel Impulse Response (CIR)}
In this section, we will briefly introduce an existing non-stationary wideband GBSM \cite{b4}. The channel is modeled as an $M_R \times M_T$ complex matrix $\mathbf{H}^G(t,\tau) = [h^G_{kl}(t,\tau)]_{M_R \times M_T}$, where $k = 1, 2, \cdots ,M_R$ and $l = 1, 2, \cdots , M_T$.
\begin{equation}
\mathbf{H}^G(t,\tau)=\sum_{n=1}^{N_{\mathrm{total}}} \mathbf{H}^G_n(t) \delta(\tau-\tau_n).
\end{equation}
%
%\noindent where $\mathbf{H}^{BL}_n(t)$ denotes the line-of-sight (LOS) part of the channel within $\text{Cluster}_n$ and $\mathbf{H}^{BN}_n(t)$ denotes the non-line-of-sight (NLOS) part.

\noindent The elements of the channel matrix $\mathbf{H}^G(t,\tau)$ are modeled as
\begin{equation}
[\mathbf{H}^G(t,\tau)]_{kl}=h^G_{kl}(t,\tau)=\sum_{n=1}^{N_{\mathrm{total}}} h^G_{kl,n}(t) \delta(\tau-\tau_n)
\end{equation}
where the channel gain $h^G_{kl,n}(t)$ of $\mathrm{Cluster}_n$ is modeled as the sum of the line-of-sight (LOS) component and the non-line-of-sight (NLOS) components.\\

--if $\mathrm{Cluster}_n \in \left\{ C_l^T \bigcap C_k^R \right\}$,
\begin{align}
h^G_{kl,n}(t) =& \underbrace{\delta(n-1)\sqrt{\frac{K}{K+1}} e^{j(2\pi f_{kl}^{\text{LOS}}t+\varphi_{kl}^{\text{LOS}})}}_{\text{LOS}} \notag\\
&+ \underbrace{\sqrt{\frac{P_n}{K+1}} \lim\limits_{S \rightarrow \infty} \frac{1}{\sqrt{S}} \sum_{i=1}^{S} e^{j(2\pi f_{n,i}t + \varphi_{kl,n,i})}}_{\text{NLOS}}
\end{align}
--if $\mathrm{Cluster}_n \notin \left\{ C_l^T \bigcap C_k^R \right\}$,
\begin{equation}
h^G_{kl,n}(t) = 0.
\end{equation}

\noindent The detailed calculations of parameters $f_{kl}^{\text{LOS}}, \varphi_{kl}^{\text{LOS}}, f_{n,i},$ and $\varphi_{kl,n,i}$ and the space-time non-stationarity of the existing GBSM can be found in \cite{b4}, and we will not go into details.

%The appearance and disappearance of clusters on the array axis represents the space-time non-stationarity of the existing GBSM \cite{b4}. Based on the birth-death process on the array axis, cluster sets for each antenna are generated. The model also incorporates the birth-death processes on the time axis \cite{b12}. The geometry among the transmitter, receiver, and clusters is updated over time. In addition to geometrical relationships, cluster sets of antennas also evolve over time and are modeled as a birth-death process.

\section{A Novel Massive MIMO BDCM}
\begin{table*}[bt]
	\caption{KEY PARAMETER DEFINITIONS OF THE NOVEL MASSIVE MIMO BDCM.}
	\begin{center}
		\begin{tabular}{|c|c|}
			\hline
			\textbf{Parameter} & \textbf{Definition}\\
			\hline
			$\text{Ant}_l^T(\text{Ant}_k^R)$ & transmit antenna $l$ (receive antenna $k$)\\
			\hline
			$\theta_{m}$ & the $m$-th virtual angle sampling NLOS AoAs\\
			\hline
			$f_{m}$ & the Doppler frequency of cluster $m$\\
			\hline
			$\varphi_{kl,m,i}$ & the phase of cluster $m$ between transmit antenna $l$ and receive antenna $k$ via the path $i$\\
			\hline
			$D_{lm,i}^T(D_{km,i}^R)$ & the distance between transmit antenna $l$ (receive antenna $k$) and cluster $m$ via the path $i$\\
			\hline
			$D_{m,i}^T(D_{m,i}^R)$ & the distance between transmit (receive) antenna center and cluster $m$ via the path $i$\\
			\hline
			$f_{kl}^{BL}$ & the LOS Doppler frequency from transmit antenna $l$ to receive antenna $k$\\
			\hline
			$\varphi_{kl}^{BL}$ & the phase of the LOS path from transmit antenna $l$ to receive antenna $k$\\
			\hline
			$\alpha_l^{BL}$ & the LOS angle of arrival (AoA) from transmit antenna $l$ to receive antenna center\\
			\hline
			$D_{kl}^{BL}$ & the LOS Distance from transmit antenna $l$ to receive antenna $k$ \\
			\hline
			$D_l^{BL}$ & the LOS Distance from transmit antenna $l$ to receive antenna center\\
			\hline
			$\alpha_v$ & the angle of the mobile station (MS) velocity\\
			\hline
		\end{tabular}
		\label{tab1}
	\end{center}
\end{table*}

The complexity of the existing GBSM is relatively high, which is not suitable for information theory and signal processing research. In view of that, in this section, we propose a novel BDCM to transfer the channel from the antenna domain into beam domain. The key parameter definitions of the novel BDCM are summarized in Table I.
%\begin{figure}[tb]
%	\centerline{\includegraphics[width=0.5\textwidth]{Fig2_Beam.pdf}}
%	\caption{Illustration of the beam domain channel.}
%	\label{fig}
%\end{figure}
\subsection{The NLOS Condition}
Illustration of the NLOS channel for the novel BDCM is shown in Fig. 1. First, let us sample the AoA of the NLOS channel uniformly distributed in the interval $[-\pi, \pi]$, and achieve the virtual angles $\theta_m$ as
\begin{equation}
\theta_m\triangleq-\pi+\frac{2\pi m}{M}, \quad m=1,\cdots, M
\end{equation}
\noindent where $M$ is the sampling number of the virtual angles. For the simplicity of the derivation, we assume the sampling number of the virtual angles $M$ is finite, and one direction with a virtual angle $\theta_m$ is represented as a beam. The AoD $\phi_m$ is dependent on the AoA, which is defined as $\phi_m=f(\theta_m)$ \cite{b4}.

The NLOS parameters are all dependent on the arrival angles, which are defined as
\begin{align}
f_m =& f_{\max } \cos \left(\theta_m-\alpha_{v}\right)\\
\varphi_{k l, m,i} =& \varphi_{0}+\frac{2 \pi}{\lambda}\left[D_{l m,i}^{T}+D_{k m,i}^{R}\right].
\end{align}
\begin{figure}[tb]
	\centering
	\includegraphics[width=0.5\textwidth]{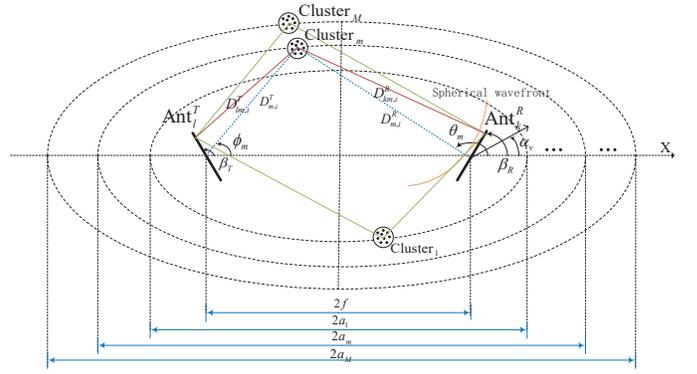}
	\caption{Illustration of the NLOS part for the novel BDCM.}
	\label{fig.}
\end{figure}
\noindent Here, the departure Angles and the arrival Angles of the paths within beam $m$ are approximated as $\phi_{m,i}\approx\phi_m$ and $\theta_{m,i}\approx\theta_m$, respectively. Moreover, we have
\begin{align}
D_{lm,i}^T =& \Big[ (D_{m,i}^T)^2 + \left(\frac{M_T-2l+1}{2}\delta_T \right)^2 \notag\\
&- D_{m,i}^T(M_T-2l+1)\delta_T\cos(\beta_T-\phi_m) \Big] \\
D_{km,i}^R =& \Big[ (D_{m,i}^R)^2 + \left(\frac{M_R-2k+1}{2}\delta_R \right)^2 \notag\\
&- D_{m,i}^R(M_R-2k+1)\delta_R\cos(\theta_m-\beta_R) \Big]
\end{align}
\noindent where
\begin{align}
D_{m,i}^T &= \frac{2a_m\sin\phi_{m,i}}{\sin\phi_{m,i}+\sin\theta_{m,i}} \\
D_{m,i}^R &= \frac{2a_m\sin\theta_{m,i}}{\sin\phi_{m,i}+\sin\theta_{m,i}}.
\end{align}

\subsection{The LOS Condition}
Similar but not the same with the existing GBSM, the LOS parameters of the novel BDCM are defined as
\begin{align}
f_{kl}^{BL} =& f_{\max } \cos \Bigg(\beta_R-\alpha_v \notag\\
&+\arcsin \bigg[\frac{D_{l}^{BL}}{D_{k l}^{BL}}  \sin \left(\alpha_{l}^{BL}-\beta_{R}\right)\bigg]\Bigg)\\
\varphi_{kl}^{BL} =& \varphi_0 + \frac{2\pi}{\lambda} D_{kl}^{\text{BL}}
\end{align}
\noindent where
\begin{align}
D_l^{BL} =& \Big[ (2f)^2+(\frac{M_T-2l+1}{2}\delta_T)^2\notag\\
&-2f(M_T-2l+1)\delta_T\cos\beta_T \Big]^{\frac{1}{2}} \\
\alpha_l^{BL} =& \arcsin\left[ \frac{(M_T-2l+1)}{2D_l^{BL}}\sin\beta_T \right]\\
D_{kl}^{BL} =& \Bigg[ (D_l^{BL})^2 + \left( \frac{M_R-2k+1}{2}\delta_R \right)^2 \notag\\
&-(M_R-2k+1)\delta_R D_l^{BL}\cos(\alpha_l^{BL}-\beta_R) \Bigg]^{1/2}.
\end{align}
%\begin{figure}[tb]
%	\centerline{\includegraphics[width=0.5\textwidth]{Fig2_Doppler.eps}}
%	\caption{The LOS Doppler frequencies of both models ($M_R = 32, M_T$ = 32, t = 1 s, $a_1$ = 100 m, f = 80 m, $\beta_R = \beta_T = \pi/2, \lambda$ = 0.12 m, $f_{\text{max}}$ = 33.33 Hz).}
%	\label{fig}
%\end{figure}
%
%As shown in Fig. 2, the LOS Doppler frequency of the proposed novel BDCM varies with the angle of MS velocity, but is not the case for the existing GBSM.
\begin{figure}[tb]
	\centering
	\includegraphics[width=0.4\textwidth]{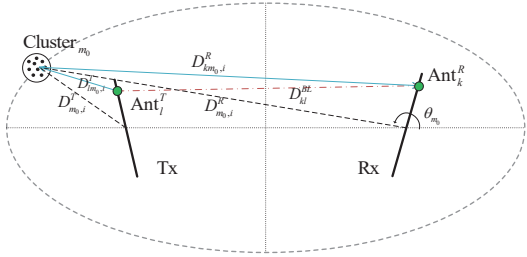}
	\caption{The modeling method of the LOS channel.}
	\label{fig.}
\end{figure}

The LOS path can be modeled also by utilizing the NLOS paths with the same or similar arrival angles $\theta_{m_0}$, which is depicted in Fig. 3, where $m_0 \in \mathcal{M}=\{1,2,\cdots,M\}$. In this case, the AoA is $\phi_{m_0}=f(\theta_{m_0})$ \cite{b4}, and we have the following approximation as
\begin{align}
D_{kl}^{BL} \approx D_{km_0,i}^R-D_{lm_0,i}^T.
\end{align}
%\begin{figure}[tb]
%	\centerline{\includegraphics[width=0.7\textwidth]{Fig3_LOS.pdf}}
%	\caption{Illustration of LOS beam domain channel.}
%	\label{fig}
%\end{figure}
%
%\noindent Thus, we have
%\begin{align}
%f_{kl}^{\text{BL}} &= f_{\max} \cos(\theta_{nm_o}-\alpha_v)\\
%\varphi_{kl}^{\text{BL}} &= \varphi_0 + \frac{2\pi}{\lambda} \left[ D_{kn}^R(\theta_{m_0})-D_{ln}^T(\theta_{m_0})\right].
%\end{align}
\subsection{The Whole Channel Model}
Based on the virtual angles, we introduce the response vectors as
\begin{align}
\mathbf{e}_T(\theta_m) &= \left[e_T(1;\theta_m), \cdots, e_T(M_T;\theta_m) \right]^T \\
\mathbf{e}_R(\theta_m) &= \left[e_R(1;\theta_m), \cdots, e_R(M_R;\theta_m) \right]^T 	
\end{align}
\noindent where
\begin{align}
e_T(l;\theta_m) &\triangleq e^{j\frac{2\pi}{\lambda}\left(D_{m,i}^T-D_{lm,i}^T\right)}, \quad l=1,\cdots, M_T \\
e_R(k;\theta_m) &\triangleq e^{j\frac{2\pi}{\lambda}\left(D_{km,i}^R-D_{m,i}^R\right)}, \quad k=1,\cdots, M_R.
\end{align}
\noindent Thus, we have
\begin{align}
h^B_{kl,n}(t) =& \delta(n-1) \sqrt{\frac{K}{K+1}} e^{ j\left[ 2\pi f_{kl}^{\text{BL}}t +\varphi_0 + \frac{2\pi}{\lambda} \left( D_{m_0,i}^R-D_{m_0,i}^T\right)  \right] } \notag\\
&\times e_R(k;\theta_{m_0}) \cdot e_T^*(l;\theta_{m_0}) + \sqrt{\frac{P_n}{(K+1)M}} \notag\\
&\times \sum_{m=1}^{M} e^{j\left[ 2\pi f_m t + \varphi_0 + \frac{2\pi}{\lambda} \left( D_{m,i}^R+D_{m,i}^T \right) \right]} \notag\\
& \times e_R(k;\theta_m) \cdot e^*_T(l;\theta_m)
\end{align}

\noindent We define
\begin{equation}
\mathbf{H}_B(t) \triangleq \mathbf{U}_R \mathbf{\tilde{H}}_{B}(t) \mathbf{U}^H_T =\mathbf{U}_R \left[ \mathbf{\tilde{H}}_{BL}(t)+\mathbf{\tilde{H}}_{BN}(t)\right]  \mathbf{U}^H_T
\end{equation}
\noindent where $\mathbf{U}_T\triangleq\left[\mathbf{e}_T(\theta_{1}), \cdots, \mathbf{e}_T(\theta_{M}) \right] \in \mathbb{C}^{M_T \times M}$, and $\mathbf{U}_R\triangleq\left[\mathbf{e}_R(\theta_{1}), \cdots, \mathbf{e}_R(\theta_{M}) \right] \in \mathbb{C}^{M_R \times M}$. We call $\mathbf{\tilde{H}}_{B}(t)$ as the beam domain channel matrix, where $\mathbf{\tilde{H}}_{BL}(t)$ and $\mathbf{\tilde{H}}_{BN}(t)$ means the LOS part and NLOS part, respectively. Due to the dependence of the AoAs and AoDs, the beam domain matrices are all diagonal and the elements of $\mathbf{\tilde{H}}_{BL}(t)$ and $\mathbf{\tilde{H}}_{BN}(t)$ are
\begin{align}
[\mathbf{\tilde{H}}_{BL}(t)]_{m_0,m_0} =& \delta(n-1) \sqrt{\frac{K}{K+1}} \notag\\
&\times e^{ j\left[ 2\pi f_{kl}^{\text{BL}}t +\varphi_0 + \frac{2\pi}{\lambda} \left( D_{m_0,i}^R-D_{m_0,i}^T\right)  \right] } \\
[\mathbf{\tilde{H}}_{BN}(t)]_{m,m} =& \sqrt{\frac{P_n}{(K+1)M}}\notag\\
& \times e^{j\left[ 2\pi f_m t + \varphi_0 + \frac{2\pi}{\lambda} \left( D_{m,i}^R+D_{m,i}^T \right) \right]}.
\end{align}

\noindent When the sampling number $M$ becomes infinite, the novel BDCM will turn to be the existing GBSM.

\subsection{Space-time Non-stationarity Modeling}
The space-time non-stationarity of the proposed BDCM is modeled the same with the existing GBSM by using the birth-death process in the array and time axes \cite{b4}.

\subsection{Computational Complexity Analysis}
The computational complexity is compared in terms of the number of ``real operations (ROs)'' \cite{b12}. In this section, the number of ROs needed for the generation of initial channel coefficients is investigated as the metric for the computational complexity of the channel model.

\subsubsection{The Existing GBSMs}
The required number of ROs for generating the existing GBSM can be calculated as
\begin{equation}
C_{\mathrm{H}^G}=C_{\mathrm{H^{GL}}}+C_{\mathrm{H^{GN}}}+1
\end{equation}
\noindent where $C_{\mathrm{H^{GL}}}$ and $C_{\mathrm{H^{GN}}}$ represent the required numbers of ROs for LOS and NLOS components of the GBSM, respectively. The number of ROs for the LOS channel is
\begin{equation}
C_{\mathrm{H^{GL}}}=174 M_R M_T+3.
\end{equation}
\noindent The number of ROs for the NLOS channel is
\begin{equation}
C_{\mathrm{H^{GN}}}= N_{\text{total}}[(S-1)(208M_R M_T+19) + 4].
\end{equation}

\noindent According to (26), the required number of ROs for generating the existing GBSM channel coefficients is
\begin{align}
C_{\mathrm{H}^G}=&174M_R M_T+3+N_{\text{total}}\notag\\
&\times [(S-1)(208M_R M_T+19) + 4].
\end{align}

\subsubsection{The Novel BDCM}
Similar to the existing GBSM, the required number of ROs for generating the proposed BDCM can be calculated as
\begin{equation}
C_{\mathrm{H}^B}=C_{\mathrm{H^{BL}}}+C_{\mathrm{H^{BN}}}+1
\end{equation}
\noindent where $C_{\mathrm{H^{BL}}}$ and $C_{\mathrm{H^{BN}}}$ represent the required numbers of ROs for LOS and NLOS components of the BDCM, respectively. The number of ROs for the LOS channel is
%\begin{equation}
%C_{\mathrm{H^{BL}}}=3+(3+1+15+115+34)M_R M_T=168M_R M_T+3.
%\end{equation}
\begin{align}
C_{\mathrm{H^{BL}}}=3+[122(M_R+M_T)+181]M.
\end{align}
\noindent The number of ROs for the NLOS channel is
%\begin{align}
%C_{\mathrm{H^{BN}}}&=3N_{\text{total}}+N_{\text{total}} (M-1) (14+140M_RM_T+15+3)\notag\\
%&=[(140M_R M_T +32)(M-1)+3]N_{\text{total}}.
%\end{align}
\begin{align}
C_{\mathrm{H^{BN}}}=3+[122(M_R+M_T)+77]M.
\end{align}
Therefore, according to (30), the required number of ROs for generating the novel BDCM channel coefficients is
%\begin{align}
%C_{\mathrm{H}^B}=&C_{\mathrm{H^{BL}}}+C_{\mathrm{H^{BN}}}+1=168M_R M_T+3\notag\\
%&+[(140M_R M_T +32)(M-1)+3]N_{\text{total}}
%\end{align}
\begin{align}
C_{\mathrm{H}^B}=6+[244(M_R+M_T)+258]M.
\end{align}

\section{Statistical Property Analysis}
Statistical properties of the channel model are very important for evaluating its performance. In this section, we will analyze four kinds of statistical functions of above-mentioned two models in order to compare their cluster-level performances.
% spatial cross-correlation function (CCF), normalized space-time-frequency correlation function(STFCF), time auto-correlation function (ACF), and frequency correlation function (FCF).

\subsection{Statistical Properties of the Existing GBSM}
The statistical properties of the existing GBSM were studied in \cite{b4}. We will give a brief description here.

\subsubsection{The space cross-correlation function (CCF)}
In the antenna domain, the normalized space CCF $\rho^G_{kl,k'l',n}(\delta_T,\delta_R;t)$ between the link connecting $\text{Ant}_k^R$ and $\text{Ant}_l^T$ and the link connecting $\text{Ant}_{k'}^R$ and $\text{Ant}_{l'}^T$ of the $n$-th cluster at time $t$ is denoted as \cite{b4}
\begin{equation}
\rho_{kl,k'l',n}^G(\delta_T,\delta_R;t) = \mathbb{E} \left[ \frac{h^G_{kl,n}(t) [h^G_{k'l',n}(t)]^*}{|h^G_{kl,n}(t)| |[h^G_{k'l',n}(t)]^*|} \right].
\end{equation}

\subsubsection{The normalized space-time-frequency correlation function (STFCF)}
Before the introduction of STFCF, we give the time-variant channel transfer function
\begin{align}
T^G_{kl}(\omega,t) = \sum_{n=1}^{N_{\text{total}}} h^G_{kl,n}(t)e^{-j2\pi\omega\tau_n}
\end{align}
\noindent where $\omega$ is the frequency. With the uncorrelated scattering (US) assumption, the normalized STFCF for the $n$-th cluster $\rho^G_{kl,k'l',n}(\delta_T,\delta_R,\Delta \omega,\Delta t;t)$ is defined by
\begin{align}
&\rho^G_{kl,k'l',n}(\delta_T,\delta_R,\Delta \omega,\Delta t;t) \notag\\
&= \mathbb{E} \left[\frac{ [h^G_{kl,n}(t)]^* h^G_{k'l',n}(t+\Delta t) e^{j2\pi\Delta \omega\tau_n]}}{|[h^G_{kl,n}(t)]^*||h^G_{k'l',n}(t+\Delta t)|} \right].
\end{align}

\subsubsection{The time auto-correlation function (ACF)}
By setting $l=l', k=k'$ , and $\Delta\omega =0$, the STFCF in (40) reduces to the time ACF $\rho^G_{kl,n}(\Delta t;t)$, which is
\begin{align}
\rho^G_{kl,n}(\Delta t;t) &= \rho^G_{kl,k'l',n}(0,0,0,\Delta t;t) \notag\\
&= \mathbb{E}  \left[\frac{[h^G_{kl,n}(t)]^* h^G_{kl,n}(t+\Delta t)}{|[h^G_{kl,n}(t)]^*||h^G_{kl,n}(t+\Delta t)|} \right].
\end{align}

\subsubsection{The frequency correlation function (FCF)}
By letting $l=l',k=k'$, and $\Delta t=0$, the STFCF in (33) reduces to the FCF $\rho^G_{kl}(\Delta \omega;t)$, which is
\begin{align}
\rho^G_{kl}(\Delta\omega;t) &= \rho^G_{kl,k'l'}(0,0,\Delta \omega,0;t) \notag\\
&= \mathbb{E} \left[ \frac{\sum\limits_{n=1}^{N_{\text{total}}} |h^G_{kl,n}(t)|^2 e^{j2\pi\Delta\omega\tau_n}}{|[T^G_{kl}(\omega,t)]^*||T^G_{k'l'}(\omega+\Delta \omega,t+\Delta t)|} \right].
\end{align}

\subsection{Statistical Properties of the Proposed Novel BDCM}
The novel BDCM is a transformation of the existing GBSM from the antenna domain into beam domain. In this section, in order to compare with the existing GBSM, we will discuss the same kinds of statistical properties for the proposed channel model.

\subsubsection{The space CCF}
The cluster-level space CCF can be defined as
\begin{equation}
\rho^B_{kl,k'l',n}(\delta_T,\delta_R;t) = \mathbb{E} \left[ \frac{h^B_{kl,n}(t) [h^B_{k'l',n}(t)]^*}{|h^B_{kl,n}(t)| |[h^B_{k'l',n}(t)]^*|} \right].
\end{equation}

\subsubsection{The normalized STFCF}

The time-variant channel transfer function $G^B_{kl}(\omega,t)$ can be calculated by
\begin{equation}
G^B_{kl}(\omega,t) = \sum_{n=1}^{N_{\text{total}}} h^B_{kl,n}(t) e^{-j2\pi\omega\tau_n}.
\end{equation}

\noindent The normalized STFCF $\rho^B_{kl,k'l',n}(\delta_T,\delta_R,\Delta \omega,\Delta t;t)$ is represented as
\begin{align}
&\rho^B_{kl,k'l',n}(\delta_T,\delta_R,\Delta \omega,\Delta t;t) \notag\\
&= \mathbb{E} \left[\frac{ [h^B_{kl,n}(t)]^* h^B_{k'l',n}(t+\Delta t) e^{j2\pi\Delta \omega\tau_n]}}{|[h^B_{kl,n}(t)]^*||h^B_{k'l',n}(t+\Delta t)|} \right].
\end{align}

\subsubsection{The time ACF}
Similar with the above-mentioned method of GBSM, we can achieve the time ACF of BDCM $\rho^B_{kl,n}(\Delta t;t)$ of the $n$-th cluster with the time-axis cluster evolution.
\begin{align}
\rho^B_{kl,n}(\Delta t;t) &= \rho_{kl,k'l',n}(0,0,0,\Delta t;t) \notag\\
&= \mathbb{E}  \left[\frac{[h^B_{kl,n}(t)]^* h^B_{kl,n}(t+\Delta t)}{|[h^B_{kl,n}(t)]^*||h^B_{kl,n}(t+\Delta t)|} \right].
\end{align}
\noindent Obviously, the time ACF of BDCM is dependent upon the time instant $t$.

\subsubsection{The FCF}
Similar with the above-mentioned method of GBSM, we can achieve the FCF of BDCM $\rho^B_{kl}(\Delta\omega;t)$, which is given by
\begin{align}
\rho^B_{kl}(\Delta\omega;t) &= \rho^B_{kl,k'l'}(0,0,\Delta \omega,0;t) \notag\\
&= \mathbb{E} \left[ \frac{\sum\limits_{n=1}^{N_{\text{total}}} |h^B_{kl,n}(t)|^2 e^{j2\pi\Delta\omega\tau_n}}{|[G^B_{kl}(\omega,t)]^*||G^B_{k'l'}(\omega+\Delta \omega,t+\Delta t)|} \right].
\end{align}

\section{Results and Discussions}
The absolute space CCFs $|\rho_{11,21,1}(0,\delta_R;t)|$ for the receiver of both two channel models are shown in Fig. 3. Each model has its reference model, simulation model, and simulation result. It can be observed that two reference models are very close. The simulation model and simulation result of BDCM are different with its reference model, but the difference is small. Thus, the transformation of a channel from the antenna to beam domain does not change the spatial correlation property of the channel.
\begin{figure}[tb]
	\centerline{\includegraphics[width=0.5\textwidth]{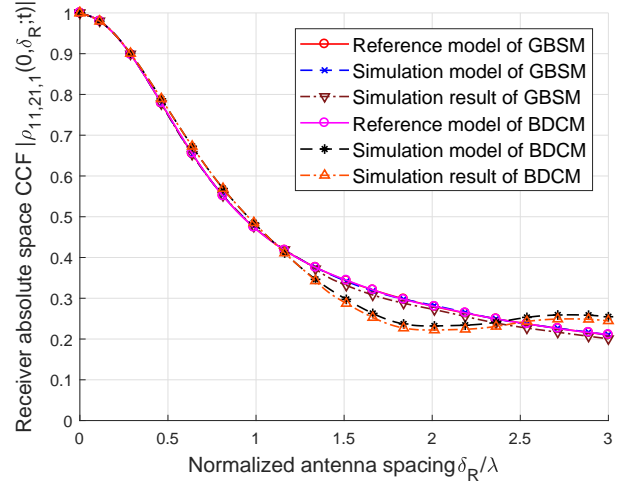}}
	\caption{Absolute receiver space CCF $|\rho_{11,21,1}(0,\delta_R;t)|$ of both models ($M_R$ = 32, $M_T$ = 32, t = 1 s, $a_1$ = 100 m, f = 80 m, $D_c^a$ = 30 m, $D_c^s$ = 50 m, $\beta_R$ = $\beta_T$ = $\pi/2$, $\lambda$ = 0.12 m, $f_{\text{max}}$ = 33.33 Hz, $\alpha_{v}$ = $\pi/6$, $\kappa$ = 5, $\bar{\alpha}_n^R = \pi/3$, $\text{NLOS}$).}
	\label{fig}
\end{figure}
	
The time correlation properties of both models are illustrated in Fig. 4. It is shown that the time ACFs of the two models are different with the time instant $t$ and the simulation represents the time non-stationarity of both models. Furthermore, the absolute value of time ACF for BDCM is very close to the GBSM. It shows that the time correlation property of the novel BDCM is almost same as the GBSM and there is no loss in time correlation property with the transformation from the antenna to beam domain.
\begin{figure}[tb]
	\centerline{\includegraphics[width=0.5\textwidth]{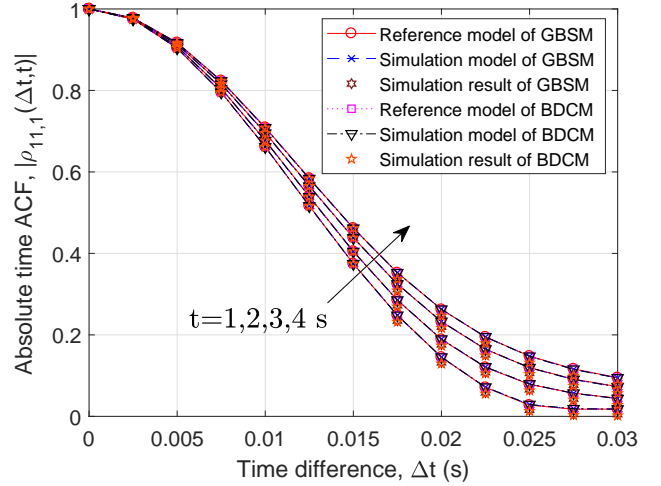}}
	\caption{Absolute time ACF of $\text{Cluster}_{1}\left|\rho_{11,1}(\Delta t ; t)\right|$ for both models in comparison among t = 1 s, t = 2 s, t = 3 s and t = 4 s ($M_{R}$= 32, $M_{T}$= 32, $a_{1}$=100 m, f=80 m, $D_{c}^{a}$=15 m, $D_{c}^{s}$=50 m, $\beta_{R}=\beta_{T}=\pi/2, \lambda$= 0.15 m, $\delta_{R}=\delta_{T}=0.5\lambda, f_{\max }$=33.33 Hz, $v_{c}$=0.5 m/s, $\mathrm{NLOS}, \lambda_{G}$=80 /m, $\lambda_{R}$=4 /m, $P_{F}$=0.3, $\kappa=5$).}
	\label{fig}
\end{figure}
	
Fig. 5 compares the FCFs of BDCM and the existing GBSM under different scenarios. It is shown that the FCFs of two models are exactly the same. Furthermore, the coherence bandwidth of BDCM and the reference ellipse model is almost the same under NLOS condition, which shows that both models have similar frequency correlation properties.
\begin{figure}[tb]
	\centerline{\includegraphics[width=0.5\textwidth]{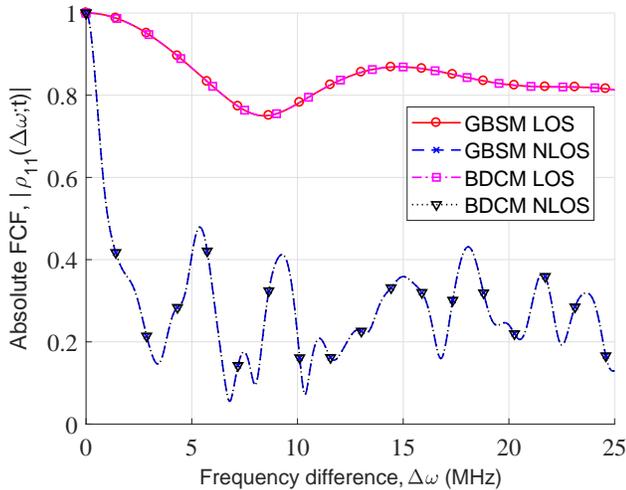}}
	\caption{Absolute FCF $|\rho_{11}(\Delta \omega; t)|$ comparison for both models between NLOS and LOS ($M_R = 32, M_T = 32, a_1$ = 100 m, f = 80 m, $D_c^a$ = 15 m, $D_c^s$ = 50 m, $\beta_R = \beta_T = \pi/2, \lambda$ = 0.15 m, $\delta_R = \delta_T$ = 0.5$\lambda, f_{\text{max}}$ = 33.33 Hz, $v_c$ = 0.5 m/s, $\lambda_G$ = 80 /m, $\lambda_R$ = 4 /m, $\eta_c$ = 0.3, $\kappa = 5$).}
	\label{fig}
\end{figure}

The complexity comparison of the existing GBSM and the proposed novel BDCM with different antenna pair numbers ($M_R \times M_T$) is shown in Fig.~6. The complexity increases linearly with the antenna pair number for the GBSM but nonlinear for the BDCM. The number of virtual angles utilized in the BDCM influences the computational complexity and the smaller the number is, the lower computational complexity the model have. It can be observed that when the number of virtual angles is equal to 1, 10 and 20 times of the total number of clusters, the computational complexity of the novel BDCM is much lower than that of GBSM. Thus, The novel BDCM is simpler than the existing GBSM, which will be more conducive to the analysis and research of signal processing and information theory.
\begin{figure}[tb]
	\centerline{\includegraphics[width=0.5\textwidth]{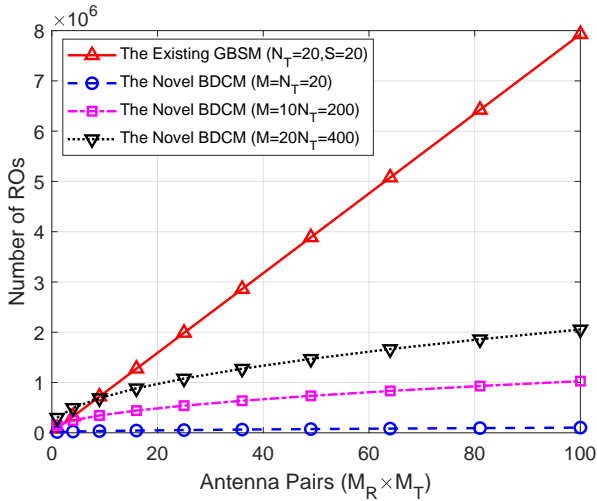}}
	\caption{Comparison of the computational complexity for both models ($M_R=M_T\in\{1,2,\cdots,10\}$, $S=20$, $N_T=20$, $M\in\{N_T, 10N_T, 20N_T\}$).}
	\label{fig}
\end{figure}

\section{Conclusions}
A novel massive MIMO BDCM with the near-field effect and space-time non-stationarity is proposed. Through the simulation results, we can find that the novel BDCM has the same cluster-level performance as the existing GBSM, and significantly reduces the computational complexity, which is more suitable for the research of signal processing and information theory. In the future work, we will further improve the novel BDCM to make it more general and applicable to various realistic scenarios, and compare it with some latest channel models and measurement data to verify the accuracy of the model, which will be helpful to information theory and signal processing researchers.

\section*{Acknowledgment}
This work was supported by the National Key R\&D Program of China under Grant 2018YFB1801101, the National Natural Science Foundation of China (NSFC) under Grant 61960206006 and 61901109, the Research Fund of National Mobile Communications Research Laboratory, Southeast University, under Grant 2020B01, the Fundamental Research Funds for the Central Universities under Grant 2242019R30001, National Postdoctoral Program for Innovative Talents (No. BX20180062), and the EU H2020 RISE TESTBED2 project under Grant 872172.

%\vspace{12pt}
%\color{red}
%IEEE conference templates contain guidance text for composing and formatting conference papers. Please ensure that all template text is removed from your conference paper prior to submission to the conference. Failure to remove the template text from your paper may result in your paper not being published.
	
\end{document}